\documentstyle[manuscript,aps,prl]{revtex}
\begin{document}
\draft
\title{Multi-Parameter Entanglement in Femtosecond Parametric Down-Conversion\\ }
\author{Mete~Atat\"{u}re,$^1$ Giovanni~Di~Giuseppe,$^2$ Matthew~D.~Shaw,$^2$
Alexander~V.~Sergienko,$^{1,2}$ Bahaa~E.~A.~Saleh,$^2$ and
Malvin~C.~Teich$^{1,2}$}
\address{Quantum Imaging Laboratory,\\
$^1$Department of Physics and $^2$Department of Electrical and Computer
Engineering,\\ Boston University, 8 Saint Mary's Street, Boston, MA 02215}

\date{\today}
\maketitle
\begin{abstract}
A theory of spontaneous parametric down-conversion, which gives
rise to a quantum state that is entangled in multiple parameters,
such as three-dimensional wavevector and polarization, allows us
to understand the unusual characteristics of fourth-order quantum
interference in many experiments, including ultrafast type-II
parametric down-conversion, the specific example illustrated in
this paper. The comprehensive approach provided here permits the
engineering of quantum states suitable for quantum information
schemes and new quantum technologies.
\end{abstract}

\pacs{42.50.Dv, 42.65.Re, 42.65.Ky, 03.67a}

Entanglement \cite{Schrodinger} is, undoubtedly, one of the most
fascinating features of quantum mechanics. Spontaneous parametric
down-conversion (SPDC) \cite{SPDC},  a nonlinear optical
phenomenon, has been one of the most widely used sources of
entangled quantum states. In this process, pairs of photons are
generated in a state that can be entangled in frequency, momentum,
and polarization when a laser beam illuminates a nonlinear optical
crystal.  The experimental arrangement for producing entangled
photon pairs is simple both in conception and in execution.

Ironically, a significant number of experimental efforts designed
to verify the nonseparability of entangled states, the hallmark of
entanglement, are carried out in the context of models that fail
to access the overall relevant Hilbert space, but rather are
restricted to only a {\em single} kind of entanglement, such as
entanglement in energy \cite{Energy}, momentum \cite{Momentum}, or
polarization \cite{Polarization}. Inconsistencies in the analysis
of down-conversion quantum-interference experiments can emerge
under such circumstances, as highlighted by the failure of the
conventional theory \cite{Femto-SPDC} of ultrafast parametric
down-conversion to characterize quantum-interference experiments
\cite{Femto-PRL}.

In this paper we present a quantum-mechanical analysis of
entangled-photon state generation via type-II SPDC, considering
{\em simultaneous} entanglement in three-dimensional wavevector
and polarization at the generation, propagation, and detection
stages. As one specific example of the applicability of this
approach, we use it to describe both new and previously obtained
\cite{Femto-PRL} results of SPDC experiments with a femtosecond
pump.  Our analysis confirms that the inconsistencies between
existing theoretical models and the observed data in femtosecond
down-conversion experiments can indeed be attributed to a failure
of considering the full Hilbert space spanned by the
simultaneously entangled quantum variables. Femtosecond SPDC
models have heretofore ignored transverse wavevector components
and have thereby not accounted for the previously demonstrated
angular spread \cite{Angle-Spread} of the down-converted light.
The approach presented here is suitable for Type-I, as well as
Type-II, spontaneous parametric down-conversion in the paraxial
approximation, which is valid for the great preponderance of
experimental SPDC efforts to date.

Our study leads to a deeper physical understanding of hyperentangled
photon states and, concomitantly, provides a route for engineering these states
for specific applications, including quantum information processing.

{\em Hyperentangled-State Generation.---}With this motivation we
present a multidimensional analysis of the entangled-photon state
generated via SPDC.  To admit a broad range of possible
experimental schemes we consider, in turn, three general and
fundamentally distinct stages in any experimental apparatus: the
generation, propagation, and detection of the quantum state
\cite{Duality}.

We begin with generation. By virtue of the weak nonlinear interaction, we consider the state generated within the confines
of first-order time-dependent
perturbation theory:

\begin{equation}
\label{Psi-Definition}
| \Psi^{(2)}\rangle ~ \sim ~{{\rm i}\over{\hbar}}\int\limits_{\,\,\, t_{0}}^{ \,\,\, t}dt'~\hat H_{\rm int}(t')~|0\rangle \,.
\end{equation}

\noindent Here $\hat H_{\rm int}(t')$ is the interaction Hamiltonian,
$(t_{0},t)$ is the duration of the interaction, and $|0\rangle$ is the
initial vacuum state.  The interaction Hamiltonian governing
this phenomenon is \cite{Wolf}

\begin{equation}
\label{Hamiltonian}
\hat H_{\rm int}(t') \sim \chi^{(2)}\int\limits_{\,\,\,\,\,\, {V}}d{\bf r}~\hat E_{p}^{(+)}(t',~{\bf r})\,\hat E_{o}^{(-)} (t',~{\bf r})\,\hat E_{e}^{(-)}(t',~{\bf r})~+~{\rm H.c.}\,,
\end{equation}

\noindent where $\chi^{(2)}$ is the second-order susceptibility
and $V$ is the volume of the nonlinear medium in which the
interaction takes place.~ The symbol $\hat E_{j}^{(\pm)}(t',~{\bf
r})$ represents the ~positive- ~~(negative-) frequency portion of
the $j$th electric-field operator, with the subscript $j$
representing the pump ($p$), ordinary ($o$), and extraordinary
($e$) waves at time $t'$ and position ${\bf r}$, and {\rm H.c.}
stands for Hermitian conjugate. In the paraxial approximation, the
polarization of each photon ($o,e$) may be assumed to be
independent of frequency and wavevector (the dependence at large
angles is considered in Ref. \cite{Migdall}). Because of the high
intensity of the pump field we take the coherent-state laser beam
to be classical, with an arbitrary spatiotemporal profile given by

\begin{equation}
\label{Pump-3DGeneral}
E_{p}({\bf r},t)=\int d{\bf k}_{p}~{\tilde E}_{p}({\bf k}_{p})e^{{\rm
i}{\bf k}_{p} \cdot {\bf r}}e^{{\rm -i}\omega_{p}({\bf k}_{p})t}\,,
\end{equation}

\noindent where ${\tilde E}_{p}({\bf k}_{p})$ is the
complex-amplitude profile of the field as a function of the
wavevector ${\bf k}_{p}$.

In contrast with previous models we consider the wavevector to be
three-dimensional, with a transverse wavevector ${\bf
  q}_{p}$ and frequency $\omega_{p}$, so that Eq. (\ref{Pump-3DGeneral}) takes the form

\begin{equation}
\label{Pump-General}
E_{p}({\bf r},t)=\int d{\bf q}_{p}\,d\omega_{p}~{\tilde
  E}_{p}({\bf q}_{p};\omega_{p})e^{{\rm i}\kappa_{p}z}e^{{\rm i}{\bf q}_{p} \cdot {\bf x}}e^{{\rm -i}\omega_{p}t}\,,
\end{equation}

\noindent where ${\bf x}$ spans the transverse plane perpendicular
to the propagation direction $z$.  In a similar way the signal and
idler fields can be expressed in terms of the quantum-mechanical
creation operators $\hat a^{\dagger}({\bf q},\omega)$ for the
$({\bf q},\omega)$ modes as

\begin{equation}
\label{Field-Generation}
\hat E_{j}^{(-)}({\bf r},t)=\int d{\bf q}_{j}\,d\omega_{j}~e^{{\rm -i}\kappa_{j}z}e^{{\rm -i}{\bf q}_{j} \cdot {\bf x}}e^{{\rm i}\omega_{j}t}\,\hat a^{\dagger}_{j}({\bf q}_{j},\omega_{j})\,,
\end{equation}

\noindent where the subscript $j=o,e$.  The longitudinal component
of ${\bf k}$, denoted $\kappa$, can be written in terms of the
$({\bf q},\omega)$ pair as \cite{Duality}

\begin{equation}
\label{Kappa-General}
\kappa=\sqrt{\left[n_{e}(\omega,\theta)\,\omega \over
c\right]^{2}-|{\bf q}|^2}\,,
\end{equation}

\noindent where $\theta$ is the angle between  ${\bf k}$ and the
optical axis of the nonlinear crystal, $n_{e}(\omega,\theta)$ is
the extraordinary index of refraction in the nonlinear medium, and
$c$ is the speed of light in vacuum.  Note that the extraordinary
refractive index, $n_{e}(\omega,\theta)$, in Eq.
(\ref{Kappa-General}) should be replaced by the ordinary
refractive index, $n_{o}(\omega)$, when calculating $\kappa$ for
ordinary waves.

Substituting Eqs. (\ref{Pump-General}) and (\ref{Field-Generation}) into Eqs.
(\ref{Psi-Definition}) and (\ref{Hamiltonian}) yields the wavefunction at the
output of the nonlinear crystal:

\begin{equation}
\label{Psi-General}
|\Psi^{(2)}\rangle \sim \int d{\bf q}_{o}d{\bf q}_{e}\,d\omega_{o}d\omega_{e} ~\Phi({\bf q}_{o},{\bf q}_{e};\omega_{o},\omega_{e})\hat a^{\dagger}_{o}({\bf q}_{o},\omega_{o})\hat a^{\dagger}_{e}({\bf q}_{e},\omega_{e})|0\rangle\,,
\end{equation}

\noindent with

\begin{equation}
\label{Phi-General}
\Phi({\bf q}_{o},{\bf q}_{e};\omega_{o},\omega_{e})~=~{\tilde E}_{p}({\bf q}_{o}+{\bf q}_{e};\omega_{o}+\omega_{e})\, L\,{\rm sinc}({{L\Delta}\over 2})e^{{\rm -i}{{L\Delta}\over2}}\,.
\end{equation}

\noindent Here $\Delta = \kappa_{p}-\kappa_{o}-\kappa_{e}$ where
$\kappa_{j}$ ($j=p, o, e$) is related to the indices $({\bf
q}_j,\omega_j)$ via relations similar to Eq.
(\ref{Kappa-General}).  The nonseparability of the function
$\Phi({\bf q}_{o},{\bf q}_{e};\omega_{o},\omega_{e})$ in Eqs.
(\ref{Psi-General}) and (\ref{Phi-General}), recalling
(\ref{Kappa-General}), is the hallmark of {\em simultaneous}
multi-parameter entanglement.

{\em Hyperentangled-State Propagation.---}Propagation between the planes of generation and detection is characterized by the classical transfer function of the optical system. The biphoton probability amplitude at the space-time coordinates $({\bf x}_{A},t_{A})$ and $({\bf x}_{B},t_{B})$, where detection will take place, is defined by \cite{Wolf},

\begin{equation}
\label{Biphoton-Definition}
A({\bf x}_{A},{\bf x}_{B};t_{A},t_{B})=\langle0| \hat E_{A}^{(+)}({\bf x}_{A},t_{A})\hat E_{B}^{(+)}({\bf x}_{B},t_{B}) |\Psi^{(2)}\rangle \,.
\end{equation}

\noindent The explicit forms of the quantum fields
present at the detection locations are represented by
\begin{eqnarray}
\label{Field-Detector-General}
\hat E_{A}^{(+)}({\bf x}_{A},t_{A})=\int  d{\bf q}\,d\omega~e^{{\rm -i}\omega
t_{A}}\left[{\cal H}_{Ae}({\bf x}_{A},{\bf q};\omega)\hat a_{e}({\bf q},\omega)+{\cal
H}_{Ao}({\bf x}_{A},{\bf q};\omega)\hat a_{o}({\bf q},\omega)\right]\,,\nonumber\\\hat E_{B}^{(+)}({\bf x}_{B},t_{B})=\int
 d{\bf q}\,d\omega~e^{{\rm -i}\omega t_{B}}\left[{\cal
H}_{Be}({\bf x}_{B},{\bf q};\omega)\hat a_{e}({\bf q},\omega)+{\cal
H}_{Bo}({\bf x}_{B},{\bf q};\omega)\hat a_{o}({\bf q},\omega)\right]\,,
\end{eqnarray}

\noindent where the transfer function ${\cal H}_{ij}$ ($i=A,B$ and $j=e,o$)
describes the propagation of a
$({\bf q},\omega)$ mode from the nonlinear-crystal output plane to the detection plane. Substituting Eqs.
(\ref{Psi-General}) and (\ref{Field-Detector-General}) into Eq. (\ref{Biphoton-Definition}) yields a general form for the biphoton probability amplitude:

\begin{eqnarray}
\label{Biphoton-General}
A({\bf x}_{A},{\bf x}_{B};t_{A},t_{B})&=\int d{\bf q}_{o}d{\bf q}_{e}\,d\omega_{o}d\omega_{e}&~\Phi({\bf q}_{o},{\bf q}_{e};\omega_{o},\omega_{e})\nonumber\\&&\times
\left[{\cal H}_{Ae}({\bf x}_{A},{\bf q}_{e};\omega_{e}){\cal H}_{Bo}({\bf x}_{B},{\bf q}_{o};\omega_{o})\,e^{{\rm
-i}(\omega_{e}t_{A}+\omega_{o}t_{B})}\right.\nonumber\\&&\left.~~~+
{\cal H}_{Ao}({\bf x}_{A},{\bf q}_{o};\omega_{o}){\cal H}_{Be}({\bf x}_{B},{\bf q}_{e};\omega_{e})\,e^{{\rm
-i}(\omega_{o}t_{A}+\omega_{e}t_{B})}\right].
\end{eqnarray}

\noindent This function can be separated into polarization-dependent and -independent terms, as necessary, for any particular configuration.  By choosing explicit forms of the functions ${\cal H}_{Ae}$, ${\cal H}_{Ao}$, ${\cal H}_{Be}$, and
${\cal H}_{Bo}$, the overall biphoton probability
amplitude can be sculpted as desired.

{\em Hyperentangled-State Detection.---}The formulation of the detection process depends on the scheme to be used. Slow detectors, for example, impart temporal integration while finite area detectors impart spatial integration. Quantum-interference experiments typically make use of just such detectors. Under these conditions, the coincidence count rate $R$ is readily expressed in terms of the biphoton probability amplitude:

\begin{equation}
\label{Coincidence-Definition}
R = \int d{\bf x}_{A}d{\bf x}_{B}\,dt_{A}dt_{B}~|A({\bf x}_{A},{\bf x}_{B};t_{A},t_{B})|^2\,.
\end{equation}

{\em Example: Quantum Interference in Ultrafast SPDC.---}We now
consider a particular example that demonstrates the validity of
our analysis: an ultrafast polarization quantum-interference
experiment of the form illustrated in Fig. 1(a). Details of the
experimental arrangement and protocol can be found in an earlier
work \cite{Femto-PRL}; in the analysis offered there we made use
of a phenomenological model that considered a collection of
contributions from different regions in the nonlinear crystal
that, in the absence of a full quantum-mechanical model, were
conjectured to be independent and distinguishable. With the help
of the general spatiotemporal quantum-mechanical approach
developed here, we are now in a position to provide a complete
analysis of those data along with new data in which filtering was
used, presented in Figs. 2 and 3, respectively.

For the polarization-interferometer arrangement illustrated in
Fig. 1(a), in the presence of a polarization-dependent relative
temporal delay $\tau$, Eq. (\ref{Field-Detector-General}) can be
conveniently separated into polarization-dependent and
-independent terms via the relation

\begin{equation}
\label{Specific-H}
{\cal H}_{ij}({\bf x}_{i},{\bf q};\omega)~=~({\bf e}_{i}\cdot {\bf e}_{j})\,e^{{\rm -i}\omega \tau \delta_{ej}}\,H({\bf x}_{i},{\bf q};\omega)\,,
\end{equation}

\noindent where $i=A,B$ and $j=e,o$. The symbol $\delta_{ej}$ is
the Kronecker delta so that $\delta_{ee}=1$ and $\delta_{eo}=0$.
The unit vector ${\bf e}_{i}$ describes the orientation of the
polarization analyzers in the experimental apparatus [see Fig.
1(a)], while ${\bf e}_{j}$ is the unit vector that describes the
polarization of the down-converted photons; the function $H({\bf
x}_{i},{\bf q};\omega)$ is the transfer function of the
polarization-independent elements of the system such as free
space, filters, apertures, and lenses, as illustrated in Fig.
1(b). The paraxial approximation satisfactorily describes our
experiments so that the explicit form of $H$ in Fig. 1(b) becomes

\begin{equation}
\label{Transfer-Function}
H({\bf x},{\bf q};\omega)=\left[e^{{\rm i}{\omega \over c}(d_{1}+d_{2}+f)}e^{{\rm
    -i}{\omega|{\bf x}|^2\over 2cf}\left[{d_{2}\over f}-1\right]}e^{{\rm
    -i}{d_{1}c\over2\omega}|{\bf q}|^2}\tilde{P}({\omega \over cf}{\bf x}-{\bf q})\right]{\cal F}(\omega)\,,
\end{equation}

\noindent where $d_1$, $d_2$, and $f$ (focal length of the lens)
are indicated, $\tilde{P}$ is the aperture function $p({\bf x})$
in the Fourier domain, and ${\cal F}(\omega)$ is the spectral
filter function.

Using Eqs. (\ref{Specific-H}) and (\ref{Transfer-Function}) in Eq. (\ref{Biphoton-General}), the biphoton probability amplitude for the arrangement shown in Fig. 1(a) therefore becomes

\begin{eqnarray}
\label{Biphoton-Ultrafast}
A({\bf x}_{A},{\bf x}_{B};t_{A},t_{B})&=\int &d{\bf q}_{o}d{\bf q}_{e}\,d\omega_{o}d\omega_{e}~\Phi({\bf q}_{o},{\bf q}_{e};\omega_{o},\omega_{e})e^{{\rm -i}\omega_{e}\tau}\nonumber\\&&\times
\left[({\bf e}_{A}\cdot {\bf e}_{e})({\bf e}_{B}\cdot
{\bf e}_{o})~H({\bf x}_{A},{\bf q}_{e};\omega_{e})H({\bf x}_{B},{\bf q}_{o};\omega_{o})e^{{\rm
-i}(\omega_{e}t_{A}+\omega_{o}t_{B})}\right.\nonumber\\&&\left.~~~+({\bf e}_{A}\cdot {\bf e}_{o})({\bf e}_{B}\cdot {\bf e}_{e})~H({\bf x}_{A},{\bf q}_{o};\omega_{o})H({\bf x}_{B},{\bf q}_{e};\omega_{e})e^{{\rm
-i}(\omega_{o}t_{A}+\omega_{e}t_{B})}\right]\,.
\end{eqnarray}

\noindent Using this form for the biphoton probability
amplitude in Eq. (\ref{Coincidence-Definition}) yields the coincidence-count
rate as a function of the polarization-dependent temporal delay $\tau$.

{\em Discussion.---}Figure 2 displays the observed normalized
coincidence rates (fourth-order quantum-interference patterns) for
0.5-, 1.5-, and 3.0-mm BBO crystals (symbols), in the absence of
spectral filtering, along with the expected theoretical curves
(solid), as a function of relative optical-path delay $\tau$. We
have treated the pump as a finite-bandwidth pulsed plane-wave, an
assumption that is valid in our experimental setup. The asymmetry
of the observed interference pattern clearly increases with
crystal thickness.

Figure 3 provides a set of data collected in a similar fashion,
but this time observed in the presence of a narrowband (9-nm)
spectral filter ${\cal F}(\omega)$, as illustrated in Fig. 1(b).
The most dramatic effect of including the filter is the
symmetrization of the quantum-interference patterns.  Since ${\bf
q}$ and $\omega$ are intrinsically linked by Eq.
(\ref{Phi-General}), the imposition of spectral filtering
restricts the allowable transverse wavevector spread. Spectral and
spatial filtering therefore have similar effects for
non-cross-spectrally pure light, such as that generated in SPDC
\cite{Angle-Spread}.

The increasing asymmetry and loss of visibility in Figs.~2 and 3
are observed with increasing crystal thickness, as the extent of
the $({\bf q},\omega)$ modes overlap less in space at the
detection plane. This decreased overlap leads to increased
distinguishability. This distinguishability is similar in nature
to the {\em spectral} distinguishability in the one-dimensional
model discussed in Ref.~\cite{Femto-SPDC}. The physical origin of
this behavior resides in the angle-dependence (hence ${\bf
q}$-dependence) of the extraordinary refractive index for the
down-converted photons [Eq.~(\ref{Kappa-General})]. Although the
phase-matching condition between the pump and the down-converted
photon pairs encompasses a large range of $({\bf q},\omega)$ modes
at the source, the combination of free-space propagation and the
small acceptance angle of the optical system leads to diffraction
of the SPDC beams, which, in turn, results in increased overlap
and therefore a decrease in distinguishability. Indeed, when the
aperture size becomes sufficiently small, the observed
quantum-interference patterns ultimately revert to those
calculated using the one-dimensional model that has traditionally
been employed.

{\em Acknowledgments.---} This work was supported by the National
Science Foundation.

\begin{figure}
\label{Setup}
\caption{(a) Schematic of the experimental setup for observation of quantum
  interference using femtosecond SPDC. (b) Detail of the path from the crystal output plane to the detector input plane.}
\label{autonum}
\end{figure}

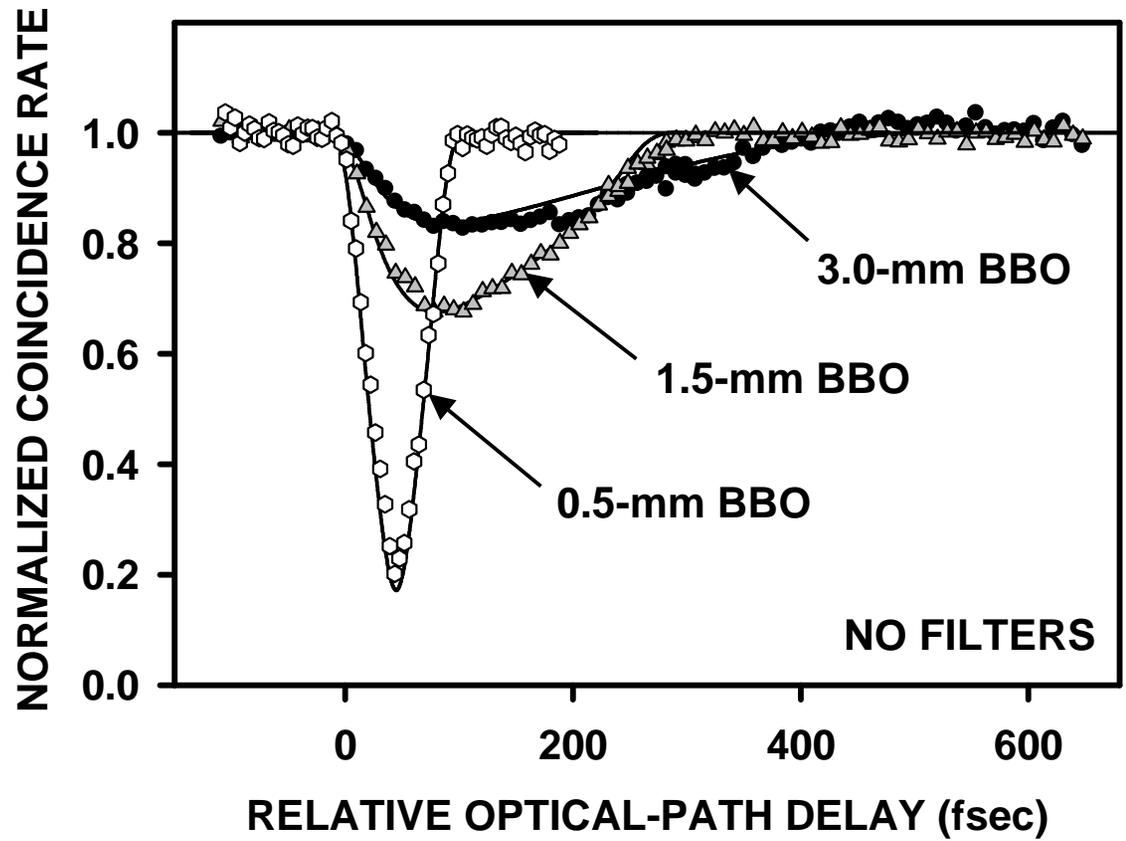
\begin{figure}
\label{Plot1}
\caption{Experimental (symbols) and theoretical (solid curves) results for the
normalized coincidence rate for BBO crystals of three different lengths
(hexagons: 0.5 mm; triangles: 1.5 mm; circles: 3.0 mm) as a function of the
relative optical-path delay $\tau$. As the crystal length increases the
fringe visibility diminishes  substantially and a dramatic asymmetry
emerges. No free parameters are used to fit the data.}
\label{autonum}
\end{figure}

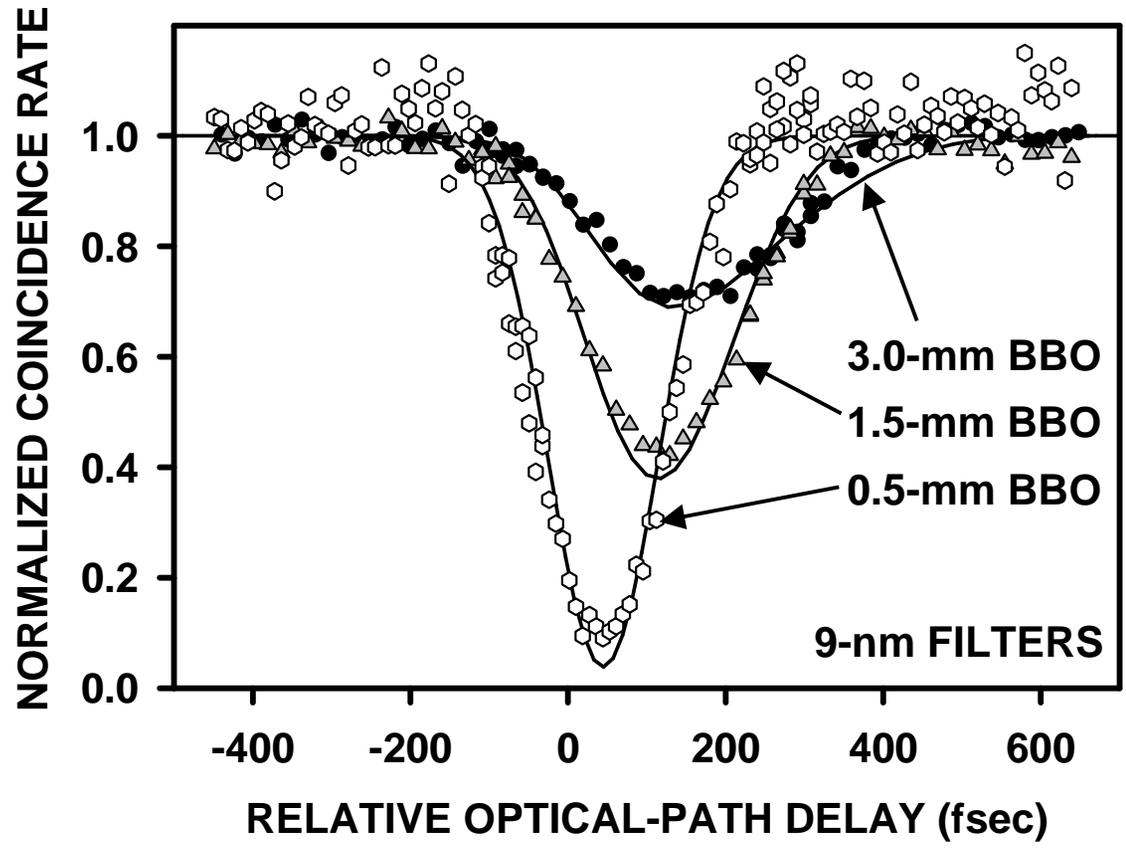
\begin{figure}
\label{Plot2}
\caption{Plots similar to those in Fig. 2 in the presence of an interference filter of 9-nm
bandwidth. The patterns are symmetrized.}
\label{autonum}
\end{figure}

\end{document}